# Epitaxy-distorted spin-orbit Mott insulator in $Sr_2IrO_4$ thin films

C. Rayan Serrao[1], Jian Liu[2,*], J.T. Heron[1], G. Singh-Bhalla[1,3], A. Yadav[1], S.J. Suresha[4], R. J. Paull[1], D. Yi[1], J.-H. Chu[1], M. Trassin[1], A. Vishwanath[2], E. Arenholz[5], C. Frontera[6], J. Železný[7], T. Jungwirth[7,8], X. Marti[1,7,9**], R. Ramesh[1,2,3,4]

[1] Department of Materials Science and Engineering, University of California, Berkeley, California 94720, USA

[2] Department of Physics, University of California, Berkeley, California 94720, USA

[3] Materials Science Division, Lawrence Berkeley National Laboratory, Berkeley, California 94720, USA.

[4] National Center for Electron Microscopy, Lawrence Berkeley National Laboratory, Berkeley, California 94720, USA.

[5] Advanced Light Source, Lawrence Berkeley National Laboratory, Berkeley, CA 94720, USA

[6] Institut de Ciència de Materials de Barcelona, ICMAB-CSIC, Campus UAB, E-08193 Bellaterra, Spain

[7] Institute of Physics ASCR, v.v.i., Cukrovarnická 10, 162 53 Praha 6, Czech Republic

[8] School of Physics and Astronomy, University of Nottingham, Nottingham NG7 2RD, United Kingdom

[9] Department of Condensed Matter Physics, Faculty of Mathematics and Physics, Charles University, Ke Karlovu 5, 12116 Praha 2, Czech Republic

**Abstract**

High quality epitaxial thin films of $J_{eff}=1/2$ Mott insulator $Sr_2IrO_4$ with increasing in-plane tensile strain have been grown on top of $SrTiO_3$(001) substrates. Increasing the in-plane tensile strain up to ~0.3% was observed to drop the c/a tetragonality by 1.2 %. X-ray absorption spectroscopy detected a strong reduction of the linear dichroism upon increasing in-plane tensile strain towards a reduced anisotropy in the local electronic structure. While the most relaxed thin film shows a consistent dependence with previously reported single crystal bulk measurements, electrical transport reveals a charge gap reduction from 200 meV down to 50 meV for the thinnest and most epitaxy-distorted film. We argue that the reduced tetragonality plays a major role in the change of the electronic structure, which is reflected in the change of the transport properties. Our work opens the possibility for exploiting epitaxial strain as a tool for both structural and functional manipulation of spin-orbit Mott systems.

* jian.liu@berkeley.edu
** xaviermarti@berkeley.edu

# Introduction

Iridium, a 5d transition metal with Z = 77, is naturally an attractive ingredient for exploiting the consequences of large spin orbit coupling (SOC) in a plethora of materials. The layer perovskite $Sr_2IrO_4$ (SIO) is no exception: the interplay of the SOC and the Coulomb repulsion, with the respective energy scales of the same order of magnitude, opens a correlated charge gap in the $t_{2g}$ band triggering an antiferromagnetic $J_{eff}=1/2$ Mott insulating state in an otherwise metallic compound [1]. The intriguing electronic and structural similarities to $La_2CuO_4$, the parent compound of high-Tc cuprates, have inspired attractive theoretical propositions including unconventional superconductivity that could emerge when suppressing the Mott insulating state in SIO [2-4]. The experimental efforts to close the Mott gap and introduce metallic conduction have been approached from the perspective of chemical substitutions, dilute doping, or high pressure experiments, mostly in single crystals, and with a spotlight on the opening of the Ir-O-Ir bonding angle. This has been corroborated by theoretical calculations showing that a 13 degrees opening of the Ir-O-Ir bonding angle should suffice for closing the gap between the upper and lower Hubbard bands in the SOC electronic structure of SIO [5]. Alternatively, the inspection of the metal-insulator transition and of the electronic structure across the $Sr_2IrO_4$, $Sr_3Ir_2O_7$, and $SrIrO_3$ family indicates that the metallic behavior is critically correlated with the increasing effective dimensionality and reduced anisotropy of these layered $J_{eff}=1/2$ systems [23].

The extensive literature on the thin film growth of many complex oxides have shown that in-plane epitaxial strains of the order of ~1% are routinely achievable. These changes significantly the c/a tetragonality of the lattice via the Poisson ratio and lead to further distortions of the unit cell geometry (both angles and lenghts), offering a natural platform for manipulating the SOC-derived physics and pursuing high temperature superconductivity. In case of the thin film SIO [6], however, a systematic study with the focus on the strain-controlled structural and electronic properties has not been performed to date.

Here we report on the epitaxial induced lattice distortion as a novel approach to perturb the local electronic structure and propose a new means to induce metallic conduction in the 5-d

SIO. We study structural and electronic properties of epitaxial thin films in which an in-plane tensile strain is induced by a SrTiO$_3$(001) substrate. We present diffraction and spectroscopic studies on a set of samples with different strains. Our work reveals that epitaxial strain induced changes in the tetragonality represent a viable path for manipulating the band gap and conduction in the SIO. We argue that the large changes observed in the electrical transport are consistent with the reduced local anisotropy in crystal and electronic structure modifications while still preserving the majority of features of the bulk Sr$_2$IrO$_4$ structure. Extrapolation of the observed reduction of activation energies with increasing distortion suggests that higher, but still feasible epitaxial strains may trigger the insulator-to-metal transition.

**Experimental Methods**

A set of epitaxial thin films with increasing in-plane tensile strain have been prepared using pulsed laser deposition. A KrF excimer laser (248 nm wavelength) was used at a repetition rate of 1 Hz. The laser beam was focused to fluency of 1.1 J/cm$^2$ on a stoichiometric Sr$_2$IrO$_4$ (SIO) target; the substrate placed at a distance of 5.5 cm. The films were deposited at a substrate temperature of 850 °C and a 1 mTorr of oxygen background pressure. At the end of the growth, the samples were cooled down in 1 atm of oxygen pressure. X-ray diffraction (XRD) analyses were carried out using a Panalytical Material Research Diffractometer. Atomic resolution high-angle annular dark-field scanning transmission electron microscopy (HAADF-STEM) images, also referred to as Z-contrast imaging was carried out using the aberration-corrected TEAM 0.5 microscope (a modified FEI Titan 80-300) operated at a voltage of 300 kV. X-ray absorption experiments were carried out on the O K-edge at 300 K by total electron yield at Beamline 4.0.2 of the Advanced Light Source in Lawrence Berkeley National Laboratory. Transport measurements were performed with a 4-point set-up, using a Keythley device (model 2400) and a cryostat by Quantum Design for temperature control.

**Results**

The high resolution STEM micrograph shown in **Fig. 1(a)** evidences the excellent epitaxial quality of our SIO thin films. The stacking of SrO and IrO$_2$ planes in ABA-ABA sequence is clearly observed across the 10 nm layer thickness, corresponding to 4 unit cells. Notice that Ir and Sr atoms are highlighted in a representative region of the image. The substrate-film interface is rather sharp and coherent. The topmost surface preserves the terraces and stepped morphology of the TiO$_2$-terminated SrTiO$_3$ substrate with a root-mean-square surface roughness below 3 Å (**Fig. 1(b)**) as expected from the intensity oscillations found in the reflection high-energy electron diffraction during the growth. Consistently, both rocking curves around the SIO(0 0 12) and SrTiO$_3$(002) have an identical full width at half maximum of 0.015 degrees, confirming the high crystallinity of the layers (**Fig. 1(d-e)**). X-ray diffraction pole figures have revealed all expected reflections consistent with a single crystal domain of Sr$_2$IrO$_4$ with the I 4$_1$/*a c d* symmetry (space group 142) and with no evident traces of spurious phases or crystal domain twinning. Azimuthal φ-scans (**Fig. 1(c)**) show a correspondence of SIO(116) and STO(101) reflections thus indicating that the epitaxial relationship is [100]SIO(001)∥[110]SrTiO$_3$(001) and confirming that the samples are single crystalline. Further confirmation is found in an exhaustive set of pole figures (data for the 60 nm thick sample are shown in **Fig. 1(f)**) which did not reveal other reflections than the ones expected for SIO(001)/ SrTiO$_3$(001). We point out the set of observed peaks shows that the conditions for the space group I4$_1$/acd are fulfilled so the doubling of the unit cell due to the alternate tilting of the IrO$_6$ octahedrons persists in our strained thin films.

Unit cell lattice parameters have been quantitatively studied by X-ray θ/2θ diffraction and reciprocal space maps. Data for three samples with increasing thickness are shown in **Fig. 2(a)**. Raw data (gray) are compared with a dynamical theory simulation (thin red line), showing a very good agreement. Reciprocal space maps around the SIO(2 0 24) reflection, shown in **Fig. 2(b-d),** reveal that the epitaxy induced in-plane tensile strains ($\varepsilon_{[100]}$) are 0.17%, 0.23% and 0.31% for the 60, 10 and 5 nm thick film, respectively. Dashed lines mark the coordinates of the bulk SIO [7-16] and solid lines indicate the location of the peak in the case of fully strained growth. Due to this in-plane tensile strain, the out-of-plane parameters contract resulting in an out-of-plane strain ($\varepsilon_{[001]}$) of -0.31%, -0.59%, and -1.40% for the above SIO film thicknesses. The unit cell volume is approximately conserved in the 60 and 10 nm thick films (V/V$_{bulk}$ = 1.000(2) and 0.999(2) respectively) and slightly reduced in the 5 nm thick film (V/V$_{bulk}$ = 0.992(2)). Hence, the structural data allow us to rule out any significant role of oxygen vacancies (which are expected to increase the unit cell volume) despite the relatively low growth pressure (1 mTorr). In summary, structural X-ray diffraction characterizations confirm spatially homogeneous high-quality thin films of SIO in which the epitaxial strain and corresponding c/a ratio can be controlled by the film thickness. In the following part of the paper we address the consequences of decreasing the film thickness on the detailed distortions of bond angles and lengths and on the electronic structure.

A detailed investigation of the bonding distortions in the 60 and 10 nm thick samples was performed by measuring an exhaustive set of relative intensities of the diffraction peaks. The refinement of the X-ray diffraction total scattering structure factors (inset in **Fig. 1(f)**) shows that apical Ir-$O_a$ bond lengths decrease from 2.14(5) Å in the 60 nm thick film to 1.96(8) Å in the 10 nm thick sample. On the other hand, the in-plane bond lengths are 2.00(4) Å and 2.00(9) Å, respectively, i.e. without a detectable change. From this we infer that the Ir-O-Ir bonding angle increases with increasing the in-plane tensile strain in the thinner SIO films. Note that the data collected for the 60 nm sample are consistent with reported bulk values of 2.151(2) Å and 1.977(6) Å for apical and basal bond lengths, respectively [14]. (From the extensive literature we selected the more reliable set of data based on high resolution neutron powder diffraction experiments.)

To investigate the resulting changes in the electronic structure, X-ray absorption experiments were carried out. To measure linear dichroism, the incident angle (with respect to the surface normal) was fixed at 60° while the linear polarization of the photon was rotated by 90° to obtain the in-plane (paralell to the surface) and out-of-plane (60° from the surface) configurations (see the inset of Fig.3). **Fig. 3** (top panel, bottom curves) shows the polarization-dependent spectra of the most relaxed, bulk-like, 60nm thick film which, as expected, show very similar lineshapes and linear dichroism to the reported bulk data [17]. In particular, the feature at the pre-edge (527 eV – 529 eV) is due to the transition into the Ir $t_{2g}$ orbitals hybridized with the 2p states of the apical oxygen ($O_a$) and planar oxygen ($O_p$). Compared with the out-of-plane configuration where this pre-peak probes the Ir $5d_{yz,zx}$-$O_p$ $2p_z$ states, the pre-peak splits into two in the in-plane configuration, representing the Ir $5d_{yz,zx}$-$O_a$ $2p_{y,x}$ and the Ir $5d_{xy}$-$O_p$ $2p_{x,y}$ states, respectively. This splitting is a result of the different Madelung potentials on $O_a$ and $O_p$ caused by the quasi-two-dimensional nature of SIO. Since the Fermi level lies in the half-filled $J_{eff} = ½$ band split off from the $t_{2g}$ manifold, this peak characterizes the $J_{eff} = ½$ holes, i.e. the upper

Hubbard band in the Mott insulating state [18]. At higher energies, the features in between 529 eV and 534 eV can be assigned to the antibonding states with the Ir $e_g$ orbitals, i.e. the wide $e_g^*$ band. Although this band is empty, the combination of the large anisotropy in the band dispersion and the $IrO_6$ octahedral elongation along the c-axis in SIO induces a strong linear dichroic effect in this region as well.

Having described the nature of these basic features in the X-ray absorption spectra, we can now compare the more strained cases with the relaxed bulk-like film. **Fig. 3** (top panel, top curves) shows the polarization-dependent spectra of the 5 nm and 10 nm films measured under the same condition. While the spectral lineshapes and features are in general agreement with that of the 60 nm film, the 5 nm film clearly has smaller dichroic effects in both the $J_{eff} = ½$ and $e_g^*$ states, indicative of the reduced anisotropy in the local electronic structure. Meanwhile, the film with the intermediate thickness of 10 nm shows an intermediate strained state in between the 60 and 5 nm samples, indicating a systematic evolution. This behavior is well reflected in the difference spectra (out-of-plane spectrum minus in-plane spectrum) of the three samples shown in Fig. 3 (bottom panel). The reduced anisotropy or the larger tensile strain can be attributed to a compensating distortion to the octahedral crystal field which is compressive in the bulk limit.

These results demonstrate the capability of the epitaxial strain to tune the orbital components of this SOC $J_{eff} = ½$ state. In particular, the compensating tetragonal distortion of the tensile strain tends to first restore the ideal composition of the $J_{eff} = ½$ state in a cubic environment and then further modify the orbital mixing in the opposite way. As suggested by the recent theoretical calculations, this effect is expected to strengthen the locking between the canting moment and the in-plane octahedral rotation [4], echoing simultaneously in magnetic and electronic transport properties. While the resulting magnetic properties will be a subject for future studies, here we address the effect on the electronic transport.

Room temperature resistivity was observed to be $\rho(300 \text{ K}) \sim 5 \times 10^{-2} \Omega \cdot cm$ for the set of thicknesses studied, with no significant variations observed with strain. In contrast, although all the samples are insulating, the temperature dependence showed a remarkable trend with strain (**Fig. 4**). In order to avoid any assumption we present the data using $\Delta \equiv 2 \cdot \frac{d}{d(1/T)} [\ln R(T)]$ which in the thermally activated transport model ($\rho \sim \exp[\Delta/(2T)]$) straightforwardly provides the thermal activation energy scale ($\Delta$) for conduction without any fitting. The raw data (inset in Fig. 4a) is collected at equally spaced temperature points. We resampled the data using linear interpolation into equally spaced points, applying a sliding average of 5 points before computing the derivative. The displayed error bars represent the maximum difference between equally spaced and interpolated data to illustrate overestimated upper bounds. Fig. 4 shows that the most relaxed 60 nm thick film (triangles) is comparable with the previously reported energy gap size obtained from optical [16] and transport measurements on SIO bulk single crystals [19] (asterisks). However, as the film thickness decreases, the corresponding in-plane tensile distortion leads to the reduction of the room temperature $\Delta$, which drops to ~150 meV in the 10 nm thick film and to ~50 meV in the 5 nm thick film. Since the intrinsic energy gap size is expected to be the only relevant activation energy scale at the highest temperature range studied, the data reveal a tunable energy gap size, between 200 meV and 50 meV in the present data set, via epitaxial strain. We note a comparison to the high-pressure single crystal experiments in which more than 30 GPa pressure is required to achieve similar changes of the bandgap [22]. We also remark that in our measurements on the 60 nm thick film the activation energy $\Delta$ monotonically decreases with decreasing temperature to ~ 20 meV at 50 K. While this behavior is similar to the temperature dependence reported in bulk single crystals, we note that it may also be understood as a cross over from thermal delocalization to the impurity states activation. Remarkably, $\Delta$ becomes temperature independent for the more strained thin films.

**Discussion**

In the remaining paragraphs we discuss the observed reduction of $\Delta$ in the context of the detailed structural analysis presented in the first part of the paper. As already mentioned in the introduction, previously reported experiments and theoretical calculations suggested that the Ir-O-Ir bonding angle ($\alpha$) plays a major role in functional properties of SIO [5,19,20]. While partial chemical substitutions [19], induced oxygen vacancies [21], or single crystal high pressure experiments [22] are expected to change the lattice geometry (bonding angles and lengths), the tensile in-plane strain of ~0.5% discussed here is also suitable to induce these unit cell distortions. Qualitatively, the increase of the boding angle $\alpha$ can be inferred from our X-ray diffraction results which show a clear strain-induced expansion of the in-plane lattice parameter while the in-plane bond-length remains approximately constant. On these grounds, tensile in-plane strain would decrease the bandgap in agreement with the density functional calculations in Ref. [5]. This conclusion is supported by the experimental findings summarized in **Fig. 3(b-c)**.

While the reduction of the energy gap with increasing in-plane tensile strain is qualitatively expected based on the above arguments, quantitative analysis reveal a surprisingly small amount of bond angle increase compared to the magnitude of energy gap reduction. The density functional calculations in Ref. [5] predict a theoretical critical value of bonding angle ~ 170° for which the upper and lower Hubbard bands are expected to merge, yet the X-ray diffraction refinement of the scattering structure factors confirms that the Ir-O1-Ir bond angle upper bound is still below 160º for the 10 nm thick sample (Table I). The small structural differences between the 10 nm and 5 nm samples suggest that the bonding angle remains still far from the theoretical critical angle for closing the gap, but the energy gap has already been reduced by a factor of four comparing to the bulk value. (Note that the direct measurement of the bond lengths and angles in the 5 nm thick film was beyond the capabilities of our present experiments.). This conclusion is supported by our corresponding density functional theory calculations [24], which show an order of a degree increase of the bonding angle α for the experimentally relevant in-plane tensile strains. We conclude that the increasing bonding angle can contribute to the observed reduction of Δ in our thin strained films, but a crucial role is also played by the ratio of the apical and planar bond lengths which drops by 1.2% between the 60 nm and 5 nm thick samples. In fact, under this rapidly increasing compression, the octahedral crystal field would stabilize or increase the weight of the planar xy orbital in the $J_{eff}$=1/2 state, favoring the in-plane transport. The shorten apical bond length would also enhance the Ir-$O_a$ hybridization, which resembles the increasing effective dimensionality during the insulator-to metal transition in the $Sr_2IrO_4$, $Sr_3Ir_2O_7$, and $SrIrO_3$ family of this layered $J_{eff}$=1/2 Mott system [23].

Finally we note that despite the visible strain-induced modulation of the absorption edge, the data in **Fig. 3** still mimic the original features of the bulk SIO structure and so does the X-ray diffraction data showing and elastic deformation which preserves the space group symmetries.

Therefore the SrTiO$_3$ substrates set the appropriate sign of the lattice-matching epitaxial strain for which the bandgap reduces while the crystal structure (symmetries, angles and distances) still resemble the bulk SIO structure. We recall that these are undismissable ingredients of the predictions for high T$_c$ superconductivity that are preserved using epitaxial strain. However, we also note that the thickness might also be a factor that affects the extraction of activation energy, since we cannot rule out the possibility that the transport properties at the surface or interface layers of the film are different from those in the bulk. An eventual missing coordination is, however, expectedly to favor correlation and a larger charge gap, which is opposite to what we observed in our electrical measurements. In addition, according to XAS data, it is unlikely that the top surface layer has a significantly different electronic structure than the bulk because missing coordination of the top Ir atoms would have strongly enhanced the linear dicrhoic effect instead. Alternatively, X-ray diffraction and the unit cell refinement evidenced a substantial tetragonality and bond-length topology change, representative of the majority diffracting volume in the sample, and in consistent trend with the expected theoretical band gap reduction. Nevertheless we consider it is worth investigating the surface and interface effect of 5d complex oxides in general in a further work.

**Conclusion**

In summary, we have demonstrated the control of the unit cell distortion in high quality epitaxial Sr$_2$IrO$_4$ thin films, achieving an up to 0.31% in-plane strain and the c/a tetragonality reduction of 1.2%. X-ray diffraction, atomic force microscopy, and high resolution scanning electron microscopy revealed flat, smooth, and correctly ABA-ABA stacked layers, with no traces of spurious phases, crystal domains and dead layers. X-ray absorption experiments probed the systematic changes of the t$_{2g}$ and e$_g$ levels as a response to the strain-induced distortion of the unit cell, indicating a reduced tetragonality and consistent with bond length obtained from the refinement of an exhaustive collection of X-ray diffraction peaks. As an alternative to the more stringent approach of chemical substitution, the epitaxial strain is presented here as a suitable tool for fine tuning the electronic structure of Sr$_2$IrO$_4$, a prototypical J$_{eff}$=1/2 SOC Mott insulator.

**Acknowledgments**

The authors are grateful to B. Dynes for fruitful comments and acknowledge V. Holy for fruitful discussions on the X-ray characterization and J. Mašek and F. Máca on theory. The authors acknowledge the support from the D.O.D - ARO MURI, E3S, and DARPA. J. T. H. acknowledges that this research was made with government support under and awarded by DOD, Air Force Office of Scientific Research, National Defense Science and Engineering Graduate (NDSEG) Fellowship, 32 CFR 168a. M. T. acknowledges the support from the NSF center for Energy Efficient Electronics Science (E3S). C.F. acknowledges financial support from Spanish Ministerio de Economía y Competitividad (Projects MAT2009-07967, Consolider NANOSELECT CSD2007-00041). X.M. acknowledges the Grant Agency of the Czech Republic No. P204/11/P339. J.Ž, J.M., and T.J. acknowledge ERC Advanced Grant 268066 and Praemium Academiae of the Academy of Sciences of the Czech Republic.

and we, therefore, ignored them in our calculation. Since magnetic ordering does not have a large effect either, we performed a non-spin-polarized calculation. We used the PBE-GGA exchange-correlation potential, R*Kmax=6.5 and mesh in the Brillouin zone composed of 1000 k points.

**Figure captions**

**Fig. 1** (a) Scanning tunnel electron microscopy image of a 10 nm thick epitaxial $Sr_2IrO_4$(001) thin film grown on top of a $SrTiO_3$(001) substrate. The image evidences the 2-D layered structure with ABA-ABA stacking of the iridium (red) and strontium (blue) oxide planes. Inset: sketch of the SIO layered structure (Ir, Sr and O are depicted by big, medium and small spheres, respectively). Bottom inset: θ/2θ scan showing only expected peaks for a c-oriented $Sr_2IrO_4$ (60 nm thick film). (b) Atomic force microscopy images showing terraces and steps with less than 1 monolayer root-mean-square roughness in a 10 nm thick samples. (c) X-ray diffraction azimuthal φ-scans indicating [110]$SrTiO_3$(001) // [100]$Sr_2IrO_4$(001) epitaxial relationship. Rocking curves around the SIO(0 0 12) and STO(002) reflections for films with thickness (d) 5 and (e) 10 nm, respectively. (f) 60 nm thick sample two dimensional projection of the reciprocal space of SIO(001)/STO(001) thin films. The abscissa $Q_{ip}$ represents the total in-plane component of the momentum transfer ($Q_{ip} = [Q_x^2 + Q_y^2]^{1/2}$) while the ordinate is the vertical component $Q_z$. Inset shows the calculated versus observed structure factors refinement for the 60 and 10 nm thick samples.Sketch illustrates the obtained Ir-octahedra distortion.

**Fig. 2** (a) X-ray diffraction θ/2θ scans for $Sr_2IrO_4$(001)/$SrTiO_3$(001) films with increasing thickness (dynamical theory diffraction simulation is shown as the red line). (b-d) Reciprocal space maps around the $Sr_2IrO_4$(2 0 24) reflection. Data indicate that the in-plane parameter monotonically contracts upon a reduction of the film thickness with a concomitant expansion of the out-of-plane parameter. Dashed red lines indicate the vertical and horizontal coordinates for bulk $Sr_2IrO_4$; the solid green line signals the in-plane parameter of the $SrTiO_3$ substrate.

**Fig. 3** (Top) Linear polarization-dependent x-ray absorption spectra at Oxygen K-edge for samples with 5, 10 and 60 nm (as indicated). (Bottom) Difference spectra corresponding to the three thicknesses. While the large linear dichroism (shown in the bottom panel) of the relaxed sample reflects the expected two-dimensional behavior, it gradually fades out due to the increasing in-plane tensile strain and the concomitant collapse of the out-of-plane parameter of the unit cell.

**Fig. 4** In-plane resistance measurements on the set of $Sr_2IrO_4$ thin films. The resistivity normalized to the room temperature value is presented in the inset. Analyses revealed a temperature dependent effective bandgap (Δ) tunable with the unit cell distortion. The most relaxed sample is mimics the reported (asterisks, adapted from Ref. 19) bulk behavior while the largest strain-induced distortion severely modifies both the slope and abscissa of the temperature dependence. Panels (b-d) plot the gap versus structural parameters. Panel e shows the theoretical increase of the Ir-O-Ir bonding angle with epitaxial strain.

**Table I**: Structural details of $Sr_2IrO_4$ (tetragonal cell, space group $I\,4_1/a\,c\,d$ no. 142). Sr and O2 (apical) occupy 16*d* (0 ¼ z) Wyckoff position, Ir occupies 8*a* (0 ¼ 3/8) Wyckoff position and O1 (basal) 16*f* (x x+¼ 1/8) position:

Table 1

|  | 60 nm film | 10 nm film |
|---|---|---|
| $a$(Å) | 5.504(8) | 5.507(8) |
| $c$(Å) | 25.710(5) | 25.639(6) |
| $z$-Sr | 0.0508(3) | 0.0476(9) |
| $x$-O1 | 0.192(8) | 0.189(17) |
| $z$-O2 | 0.9580(18) | 0.951(3) |
| $d_{Ir-O1}$(Å) | 2.00(4) | 2.00(9) |
| $d_{Ir-O2}$(Å) | 2.14(5) | 1.96(8) |
| $\theta_{O1-Ir-O1}$(°) | 153.9(9) | 153(4) |

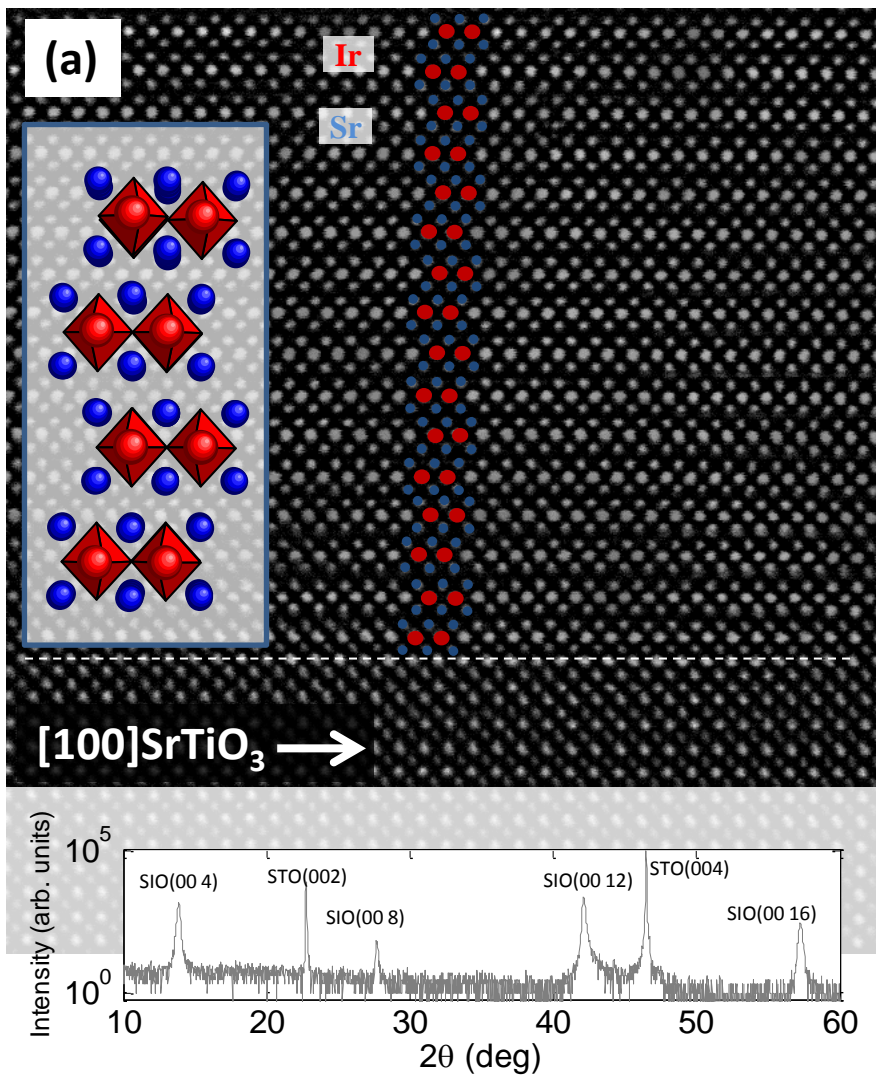
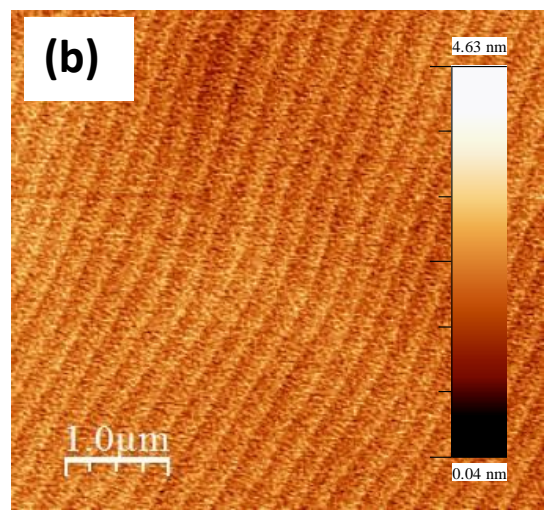
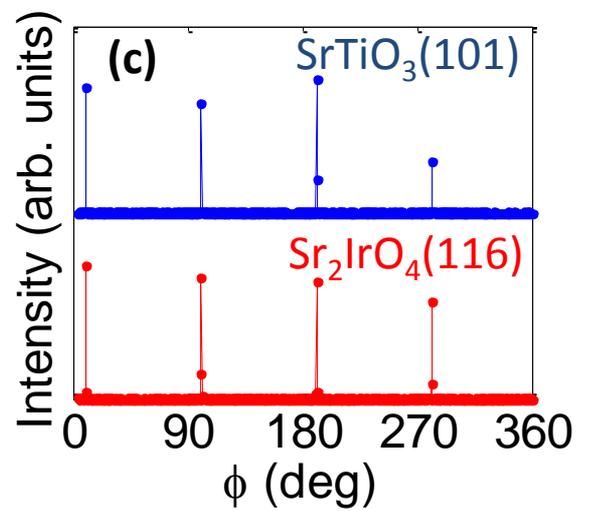
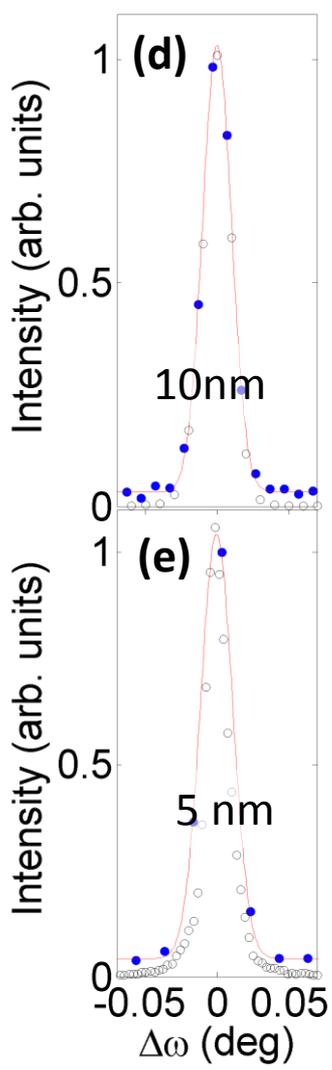
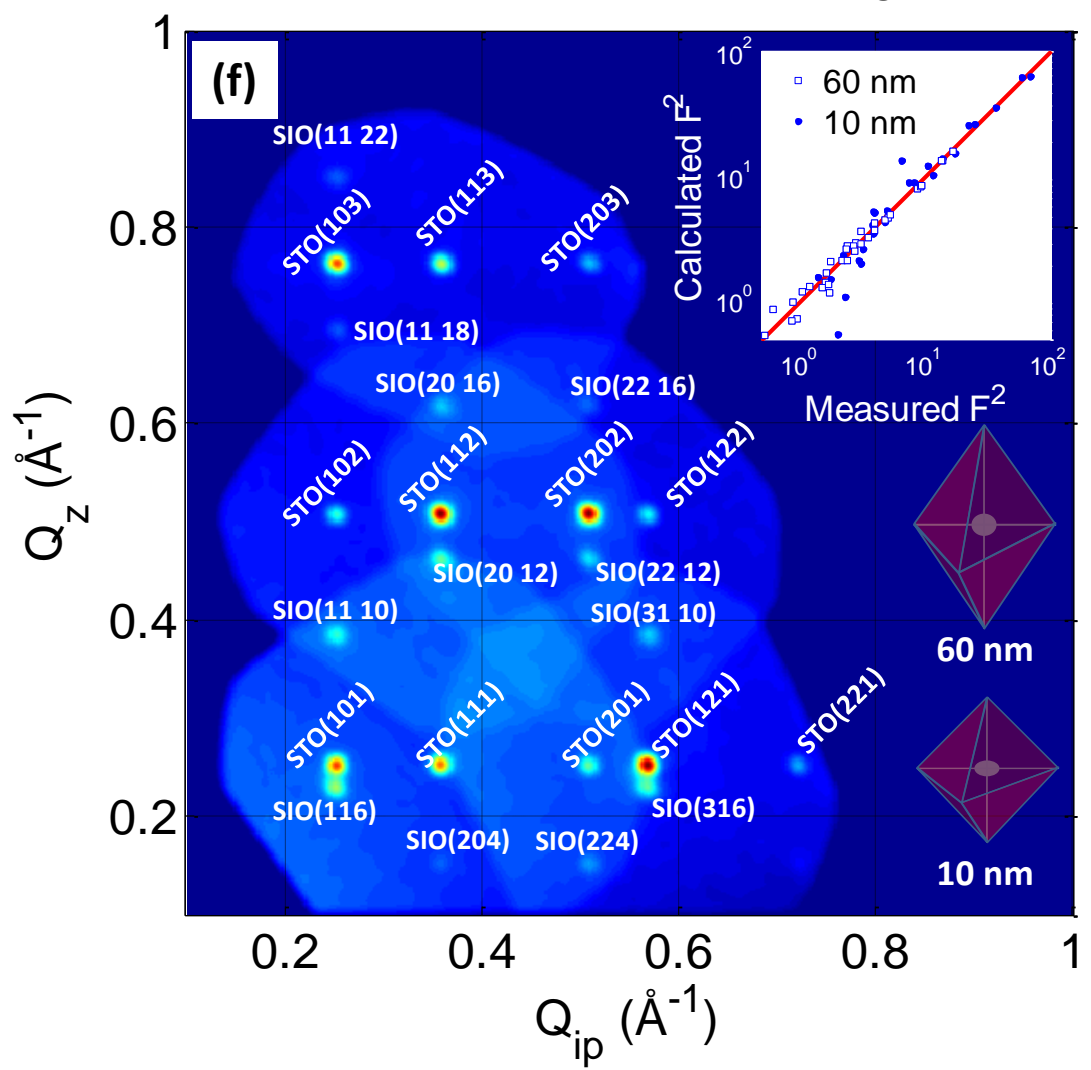

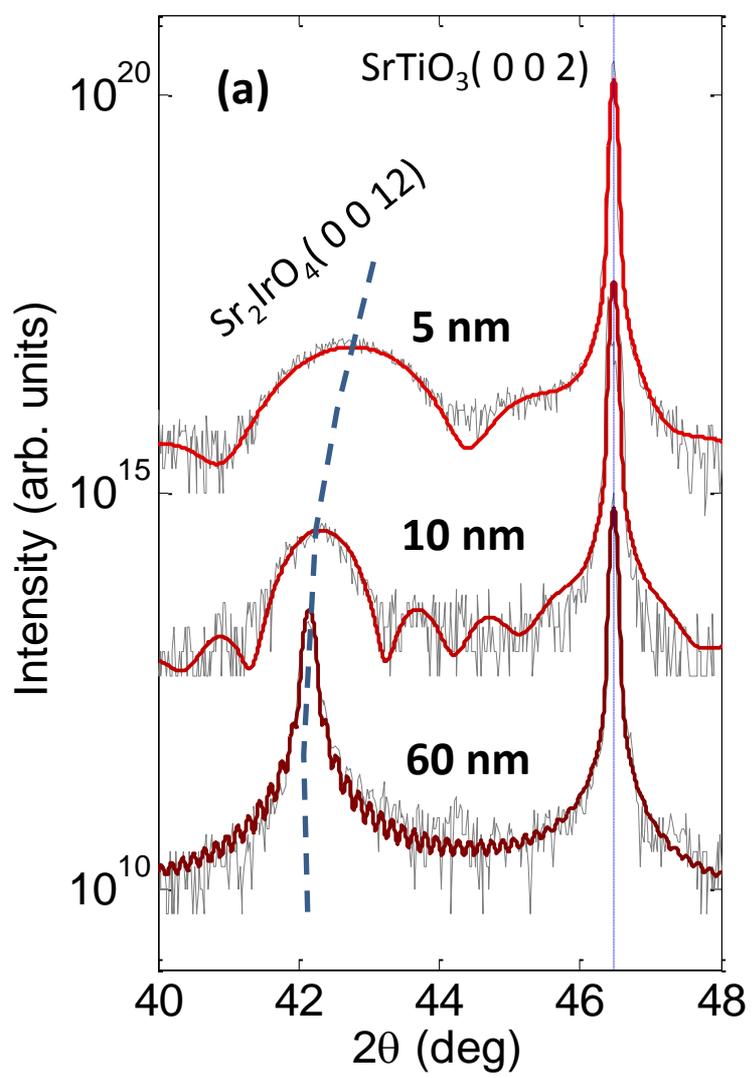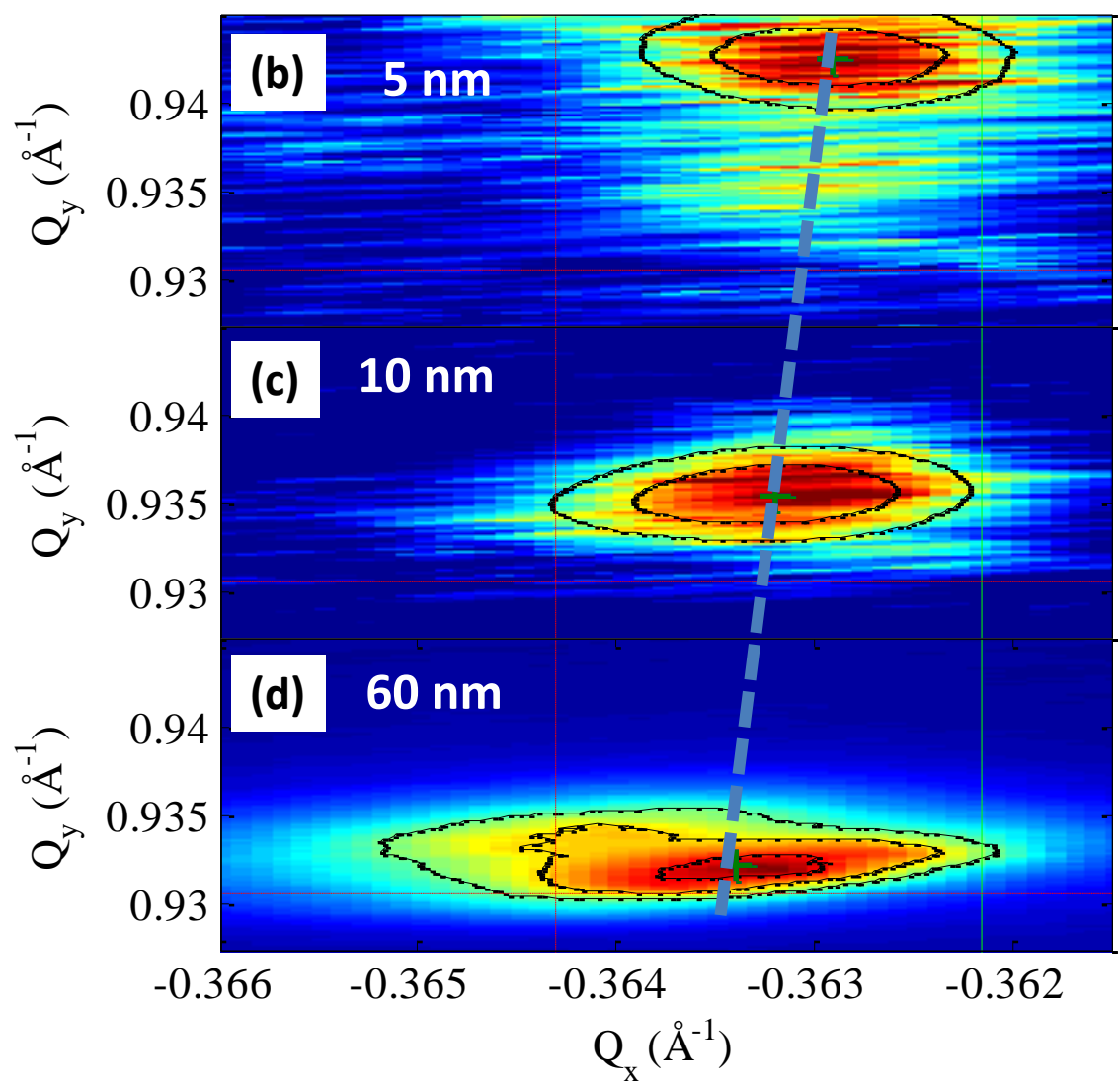

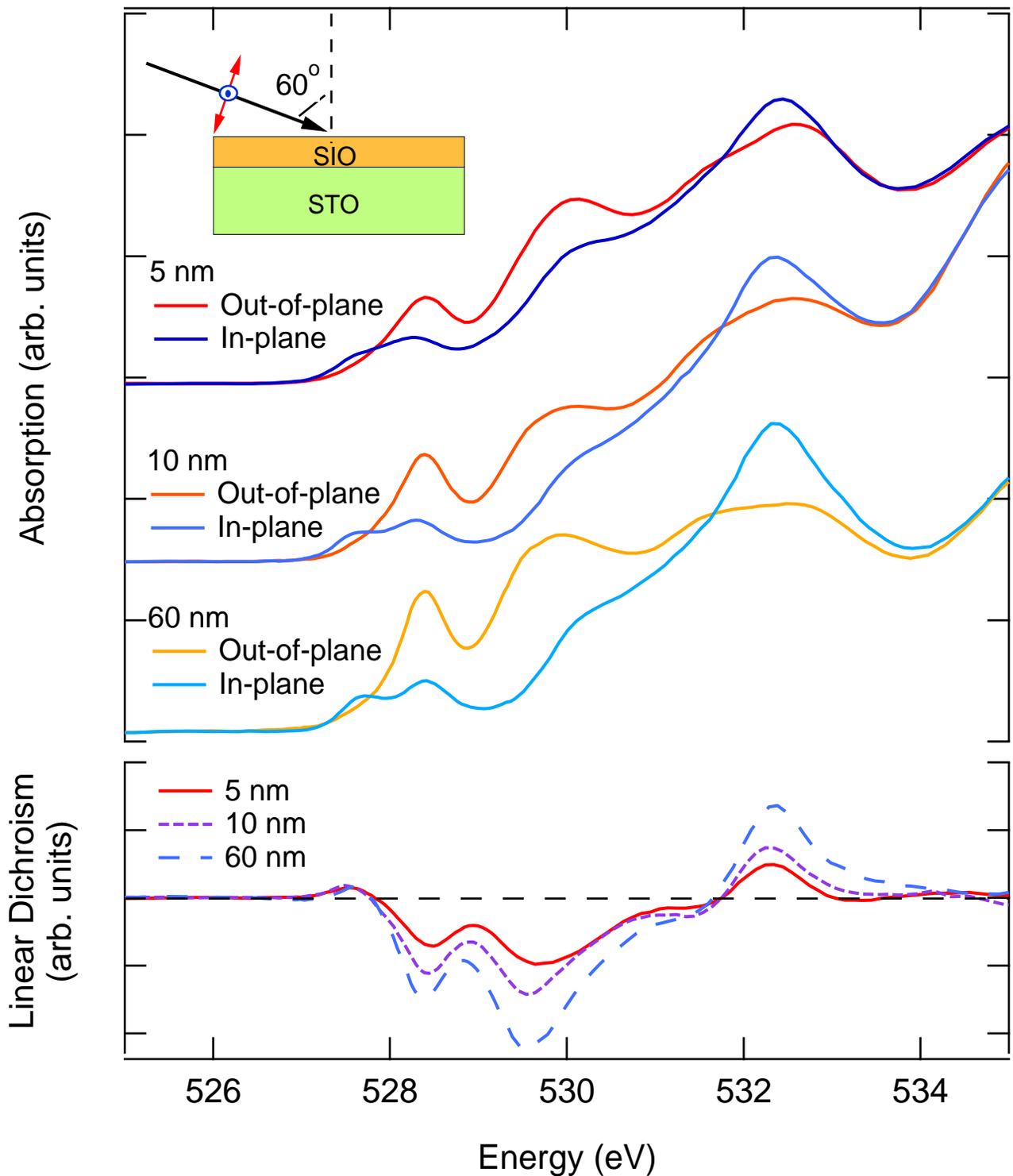

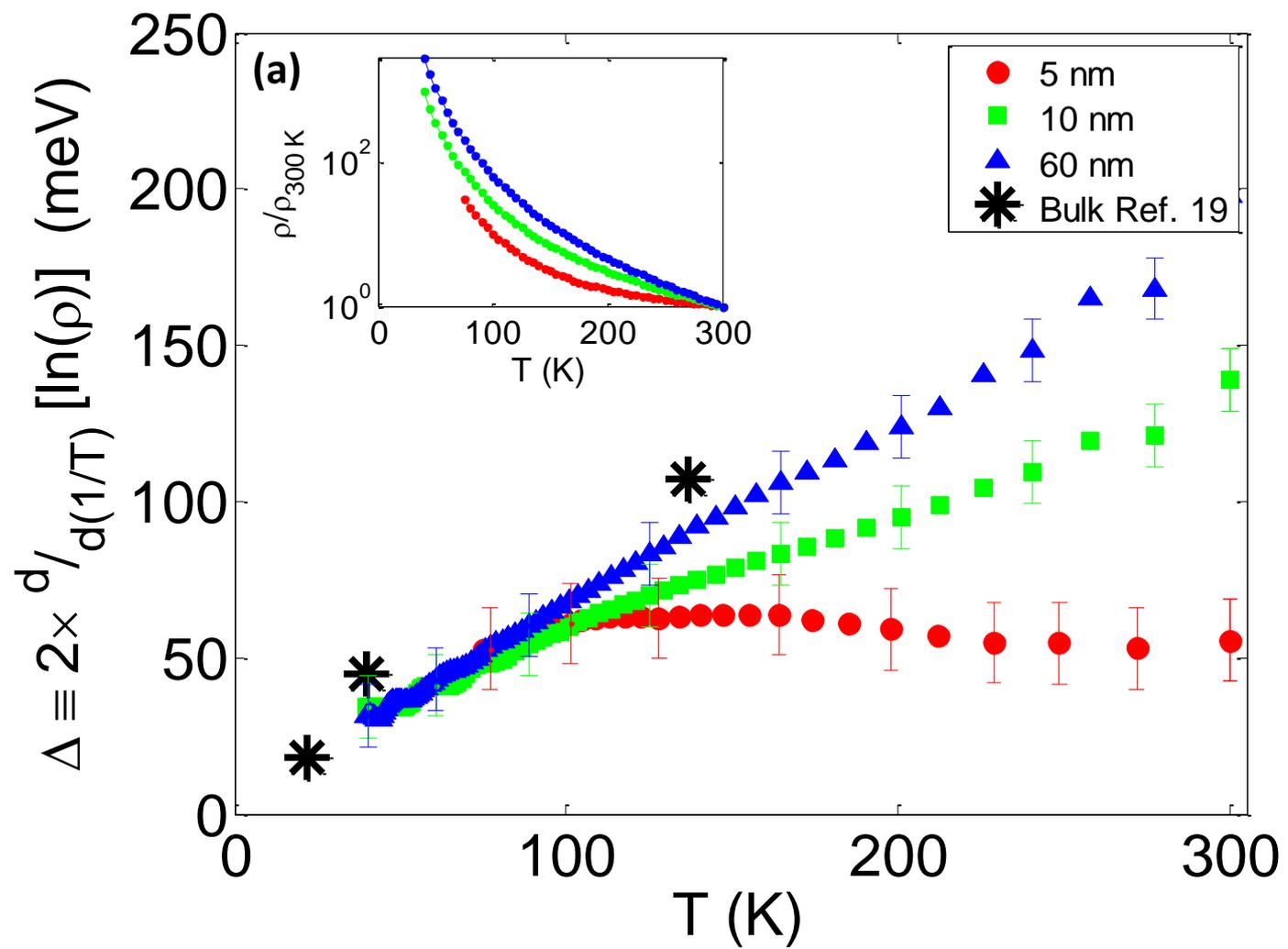

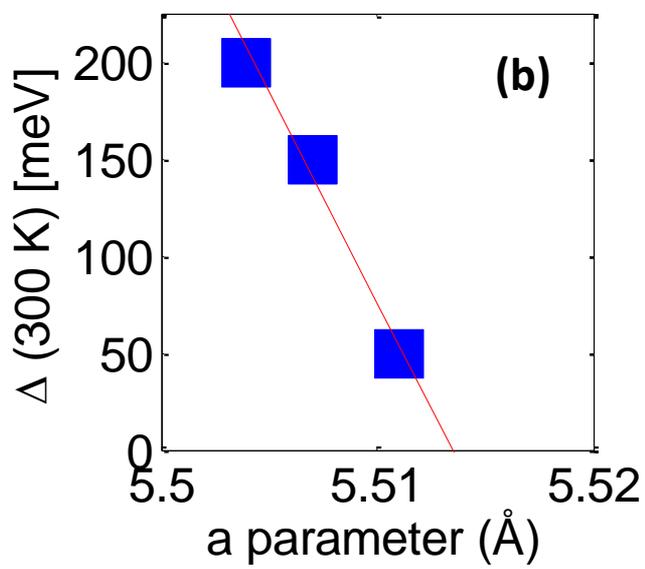
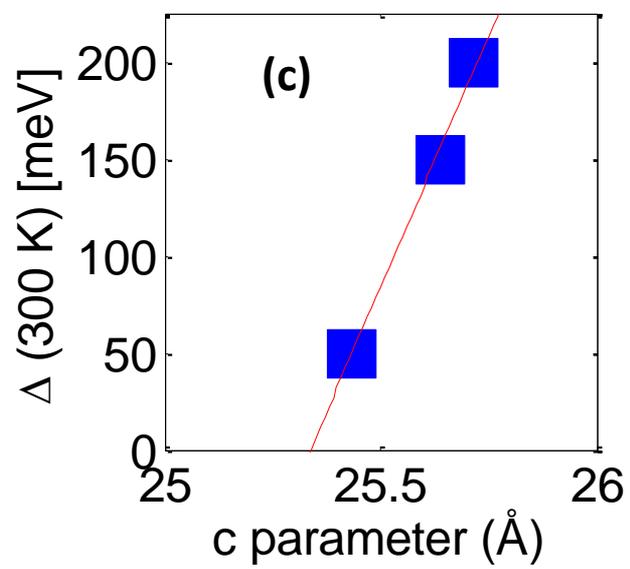

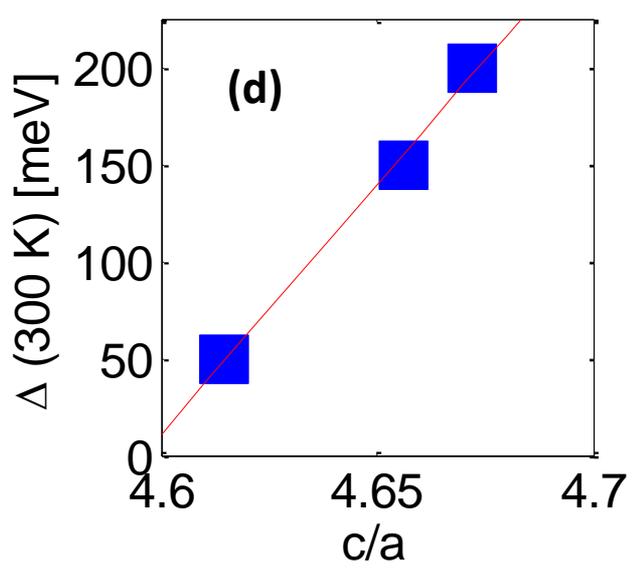
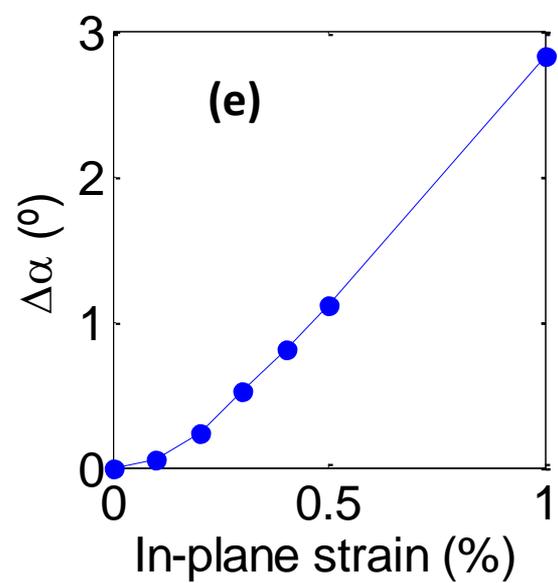